\documentclass[12pt]{article}
\usepackage{amsmath}
\usepackage{amsthm}
\usepackage{amsfonts}
\usepackage{amssymb}
\usepackage{url}
\usepackage{hyperref}
\usepackage{eucal}
\usepackage{mathrsfs}
\usepackage{graphicx,graphics}
\usepackage[T1]{fontenc}
\usepackage{url}
\usepackage{times}
\usepackage{helvet}
\usepackage{courier}
\usepackage{bm}
\usepackage{url}
\makeatletter
\renewcommand\@biblabel[1]{#1} 
\makeatother
\usepackage[english]{babel}
\usepackage{hyperref}
\newcommand\fig[1] {{\rm Figure}~\ref{fig:#1}}
\newcommand\labfig[1] {\label{fig:#1}}

\newcommand{\bfm}[1]{\mbox{\boldmath ${#1}$}}
\newcommand{\nonum}{\nonumber \\}
\newcommand\eq[1] {(\ref{#1})}
\newcommand{\beqa}{\begin{eqnarray}}
\newcommand{\eeqa}[1]{\label{#1}\end{eqnarray}}
\newcommand{\beq}{\begin{equation}}
\newcommand{\eeq}[1]{\label{#1}\end{equation}}

\newcommand{\Grad}{\nabla}
\newcommand{\Div}{\nabla \cdot}
\newcommand{\Curl}{\nabla \times}

\newcommand{\Imag}{\mathop{\rm Im}\nolimits}

\newcommand{\Md}{\partial}

\newcommand{\Ga}{\alpha}
\newcommand{\Gb}{\beta}
\newcommand{\Gd}{\delta}

\newcommand{\Gg}{\gamma}

\newcommand{\Gl}{\lambda}
\newcommand{\Gn}{\eta}

\newcommand{\Gvr}{\varrho}

\newcommand{\GU}{\Upsilon}
\newcommand{\GO}{\Omega}

\newcommand{\BGve}{\bfm\varepsilon}

\newcommand{\BGm}{\bfm\mu}

\newcommand{\CP}{{\cal P}}

\newcommand{\CT}{{\cal T}}

\newcommand{\bpm}{\begin{pmatrix}}
\newcommand{\epm}{\end{pmatrix}}

\def\b0{\bf 0}

\def\Be{{\bf e}}
\def\Bf{{\bf f}}
\def\Bg{{\bf g}}

\def\Bj{{\bf j}}
\def\Bk{{\bf k}}

\def\Br{{\bf r}}
\def\Bs{{\bf s}}
\def\Bt{{\bf t}}
\def\Bu{{\bf u}}

\def\Bx{{\bf x}}
\def\By{{\bf y}}
\def\Bz{{\bf z}}
\def\BA{{\bf A}}
\def\BB{{\bf B}}

\def\BD{{\bf D}}
 \def\BE{{\bf E}}
\def\BF{{\bf F}}
\def\BG{{\bf G}}

\def\BI{{\bf I}}

\def\BL{{\bf L}}

\def\BP{{\bf P}}
\def\BQ{{\bf Q}}
\def\BR{{\bf R}}
\def\BS{{\bf S}}
\def\BT{{\bf T}}

\def\BV{{\bf V}}
\def\BW{{\bf W}}

\begin{document}
\title{Analytic Materials}
\author{Graeme W. Milton \\ Department of Mathematics, University of Utah, USA \\ milton@math.utah.edu}
\date{\today}
\maketitle
\begin{abstract} The theory of inhomogeneous analytic materials is developed. These are 
materials where the coefficients entering the equations involve analytic functions. Three types
of analytic materials are identified. The first two types involve an integer $p$. If $p$ takes its 
maximum value then we have a complete analytic material. Otherwise it is incomplete
analytic material of rank $p$. For two-dimensional materials further progress can be made in
the identification of analytic materials by using the well-known fact that a $90^\circ$ rotation
applied to a divergence free field in a simply connected domain yields a curl-free field,  and this can then be expressed as the
gradient of a potential. Other exact results for the fields in inhomogeneous media are reviewed. Also reviewed is the subject of
metamaterials, as these materials provide a way of realizing desirable coefficients in the equations.

\end{abstract}
\section{Introduction}
\setcounter{equation}{0}

Nearly all linear equations of physics (including static equations, heat and diffusion equations, and wave equations) 
can be written in the form of a system of second order partial differential equations.
\beq \Div\left[\BL(\Bx)\Grad\Bu(\Bx)\right]+\BB(\Bx)\cdot\Grad\Bu(\Bx)
+\BA(\Bx)\Bu(\Bx)=\Br(\Bx), \eeq{2.1}
or in component form
\beq \frac{\Md}{\Md x_j}\left[L_{j\Ga i \Gb}(\Bx)\frac{\Md u_\Gb(\Bx)}{\Md x_i}\right]
+B_{\Ga i \Gb}(\Bx)\frac{\Md u_\Gb(\Bx)}{\Md x_i}
+A_{\Ga\Gb}(\Bx)u_\Gb(\Bx)=r_{\Ga}(\Bx),
\eeq{2.2}
where $i$ and $j$ are ``space indices'' taking values from $1$ to $d$, while $\Ga$ and $\Gb$ are field indices taking values
from $1$ to $m$, and where sums over repeated indices are assumed. It could be the case that $m=1$, in which case the field $\Bu(\Bx)=u(\Bx)$ is a scalar field and we can drop the field indices.
Here we regard  $\BL(\Bx)$, $\BB(\Bx)$, and $\BA(\Bx)$  as the material parameters, $\Bu(\Bx)$ and its derivatives as the fields,
and $\Br(\Bx)$ as the source term. One can write the equations of electromagnetism, acoustics, elastodynamics, heat conduction, diffusion,
and quantum physics in this form. One point to emphasize is that $\Bx$ need not represent just
spatial variables: we could have $d=4$ where $x_4$ represents time and $(x_1, x_2, x_3)$ represents a point in three-dimensional space. Because analytic functions are involved, in the analysis presented here, the coefficients and fields
have both real and imaginary parts. However, by taking the real and imaginary parts of \eq{2.2} and expressing them in terms of the real and imaginary parts of ${\bf u}({\bf x})$, it is clear that our analysis also directly applies to certain systems of linear equations with both real coefficients and real potentials.
The parameters $x_i$ need not represent cartesian coordinates, but instead could represent curvilinear coordinates. Our results apply
to partial differential equations of the form \eq{2.2} irrespective of what physical interpretation they may have. In fact the analysis here easily extends
to higher order partial differential equations as appropriate, for example, to the treatment of vibrating plates. 
In multiparticle quantum systems of $n$ particles we may choose $d=3n$ and then $\Bx$ represents the coordinates of all $n$ particles.

In our treatment we are considering the equation \eq{2.1} very generally. We do not demand that the material parameters
$\BL(\Bx)$, $\BB(\Bx)$, and $\BA(\Bx)$ be real, as they can be complex at constant frequency for the wave equations of 
electromagnetism, acoustics, elastodynamics if there is loss, or if the frequency is complex, corresponding to growing fields. We do not demand that 
they have any symmetry properties such as $L_{j\Ga i \Gb}(\Bx)=L_{i \Gb j\Ga}(\Bx)$
that generally are a consequence of Onsager's principle and which need not hold if there is some breaking of time-reversal symmetry,
such as happens when there is a magnetic field. An example is conductivity in the presence of a magnetic field where the conductivity tensor is not
symmetric due to the presence of a magnetic field. Finally we do not demand that say the imaginary part of $L_{i \Gb j\Ga}(\Bx)$ be positive semidefinite,
in the sense that the quadratic form,
\beq f(\BE)=E_{i\Gb}[\Imag(L_{i\Gb j\Ga})]E_{j\Ga}, \eeq{2.3}
is non-negative for all matrices $\BE$ with elements $E_{i\Ga}$, $i=1,2,\ldots, d $, $\Ga=1,2,\ldots, m$,
as may happen at constant frequency if there is only loss in the system, as there could be gain.

It is also to be stressed that there is typically a lot of flexibility in writing down the equations that describe a physical system. Of course,
a second order equation, like \eq{2.1}, always can be written as a system of first order equations by introducing extra
variables. For example, by introducing $\BF=\Grad\Bu$ \eq{2.1} can be written as
\beq \underbrace{\bpm \BL(\Bx) \\ \BI \epm}_{\widetilde{\BB}(\Bx)}\cdot
\Grad\underbrace{\bpm \BF(\Bx) \\ \Bu(\Bx)\epm}_{\widetilde{\Bu}(\Bx)}
+\underbrace{\bpm \BB(\Bx)+(\Div\BL(\Bx)) & \BA(\Bx) \\ -\BI & 0 \epm}_{\widetilde{\BA}(\Bx)}
\underbrace{\bpm\BF(\Bx) \\ \Bu(\Bx)\epm}_{\widetilde{\Bu}(\Bx)}=
\underbrace{\bpm \Br(\Bx) \\ 0\epm}_{\widetilde{\Br}(\Bx)},
\eeq{2.3a}
that is still of the form \eq{2.1}, but without the second order term.

Analytic materials are materials where the material parameters $\BL(\Bx)$, $\BB(\Bx)$, and $\BA(\Bx)$ involve analytic functions,
respectively. We will see that they provide a wealth of examples where solutions to the partial differential equations \eq{2.1} can be easily
found. These then can serve as benchmarks for numerical calculations, and to gain insight into what manipulations of fields are possible.
An early example for the Schr{\"o}dinger equation was discovered by Berry \cite{Berry:1998:FNM}. He realized that with a potential $V(\Bx)=iV_0(e^{iKx_1}-1)$  
(that is obviously an analytic function of $x_1$) one could explicitly
work out the strengths of all the diffracted beams. Independently, Milton \cite{Milton:2004:RAS} (see also Sections 14.4 and
14.5 of \cite{Milton:2002:TOC}) considered the scalar wave equation
\beq -\Div [a_0+\Gd q(\Bx)]\Grad u(\Bx)=\Gl u(\Bx), \eeq{2.4}
in a periodic medium with $\Gd$ being a small constant and with $q(\Bx)$ having Fourier components $\widehat{q}(\Bk)$ vanishing in half of Fourier space, say
with $\widehat{q}(\Bk)=0$ if $k_1\leq 0$. He showed that the periodic component of the Bloch solution for $u(\Bx)$
shares the property of vanishing in half of Fourier space, and moreover the band structure is exactly the same as for a homogeneous medium, with coefficient $a_0$, except possibly for the highest bands. An example
was given of a medium with a coefficient
\beq q(\Bx)=[\cos(x_1)+i\sin(x_1)]r(x_2,x_3), \eeq{2.5}
that is clearly an analytic function of $x_1$. (Here $r(x_2,x_3)$ is an arbitrary bounded function of $x_1$ and $x_2$.) 
Perhaps closely related to this (as a dilute array of scatterers can be considered as a periodic medium) is
the result of Horsley, Artoni, and La Rocca \cite{Horsley:2015:SKK} that an inclusion will not scatter if its moduli only depend on $x_1$ and have Fourier components 
that vanish in half of Fourier space. 

In a significant advance, Horsley, King, and Philbin \cite{Horsley:2016:WPC} found that a solution of an ordinary differential equation (ODE) as a function of a real valued parameter $z$ could
be analytically continued to complex $z$. Then, one could consider a real valued parameter $\Gg$ that is a function of $\Bx$, and from an
associated trajectory $z(\Gg)$ obtain a solution to the wave equation in the $\Bx$ coordinates. 

In this paper we show that the range of analytic materials can be vastly expanded. It is to be emphasized that an analytic material need not have any
periodicity and may be only defined inside some body $\GO$ as the analytic continuation of the analytic functions could have singularities outside 
$\GO$. This paper concentrates on the general theory. Applications
to specific physical equations will be considered elsewhere.

\section{A review of some other exact solutions}
\setcounter{equation}{0}
\subsection{Equivalent media using Translations}
Suppose we can find tensor fields $\BT(\Bx)$ and $\BS(\Bx)$ such that the equation
\beq \Div\left[\BT(\Bx)\Grad\Bu(\Bx)\right]+\BS(\Bx)\cdot\Grad\Bu(\Bx)=0
\eeq{2.10}
holds for {\it all fields} $\Bu(\Bx)$. Then, assuming \eq{2.1} has been solved for $\Bu(\Bx)$ and subtracting \eq{2.10} from it, we see that $\Bu(\Bx)$
solves the equations
\beq \Div\underline{\BL}(\Bx)\Grad\Bu(\Bx)+\underline{\BB}(\Bx)\cdot\Grad\Bu(\Bx)
+\underline{\BA}(\Bx)\Bu(\Bx)=\Br(\Bx), \eeq{2.11}
in a translated medium with moduli
\beq \underline{\BL}(\Bx)=\BL(\Bx)-\BT(\Bx),\quad\underline{\BB}(\Bx)=\BB(\Bx)-\BS(\Bx),\quad \underline{\BA}(\Bx)=\BA(\Bx).
\eeq{2.12}
The medium is called a translated medium because \eq{2.12} corresponds to a translation of the moduli in tensor space, at each point $\Bx$. 

Writing out \eq{2.10} in components we obtain
\beq T_{j\Ga i \Gb}(\Bx)\frac{\Md^2 u_\Gb(\Bx)}{\Md x_i\Md x_j}+\frac{T_{j\Ga i \Gb}(\Bx)}{\Md x_j}\frac{\Md u_\Gb(\Bx)}{\Md x_i}
+S_{\Ga i \Gb}(\Bx)\frac{\Md u_\Gb(\Bx)}{\Md x_i}=0.
\eeq{2.13}
Clearly, this is satisfied for all fields $\Bu(\Bx)$ if and only if
\beq T_{j\Ga i \Gb}(\Bx)=-T_{i\Ga j \Gb}(\Bx),\quad S_{\Ga i \Gb}(\Bx)=-\frac{\Md T_{j\Ga i \Gb}(\Bx)}{\Md x_j}.
\eeq{2.14}
Note, in particular, that we could choose $\BS(\Bx)=\BB(\Bx)$ and then use \eq{2.14} to determine (non-uniquely) $\BT(\Bx)$. In the setting
of \eq{2.1}, it is apparent that
we can assume $\BB(\Bx)=0$, by, if necessary, translating to an equivalent medium where this holds. This observation
is what allowed Norris and Shuvalov \cite{Norris:2011:ECT} (see also \cite{Vasquez:2013:SRP}) to eliminate such couplings
when considering transformation elastodynamics.
 
It may be the case that the moduli $\BL$, $\BB$, and $\BA$  are independent of $\Bx$, and that consequently \eq{2.1} is easily solved
for the field $\Bu(\Bx)$ when the source term $\Br(\Bx)$ and the boundary conditions are specified. Then, using translations, we obtain solutions for
the field $\Bu(\Bx)$ in a wide variety of inhomogeneous media.

Translations and equivalent media have played an important role in the theory of composites. See, in particular, Chapters 3, 4, 24, and 25
of \cite{Milton:2002:TOC} and references therein. Notably, Lurie and Cherkaev \cite{Lurie:1982:AEC,Lurie:1984:EEC}
and Murat and Tartar \cite{Murat:1985:CVH,Tartar:1985:EFC}
found that elementary bounds on the effective tensor of the "translated
medium" could produce useful, and in many cases sharp, bounds on the effective tensor of the original medium.

In the case where the equations \eq{2.1} are the Euler--Lagrange equations of an associated "energy function", the quadratic form associated
with a relevant translation usually corresponds to a null-Lagrangian: a functional whose associated Euler-Lagrange equation vanishes identically.
There is a large literature on null-Lagrangians--see, for example,
\cite{Murat:1978:CPC,Murat:1981:CPC,Murat:1987:SCC,Ball:1981:NLW,Pedregal:1989:WCW}.

\subsection{Equivalent media using coordinate transformations}

There is of course an enormous literature on the exact solution of these equations in various coordinate systems when
the coefficients $\BL(\Bx)$, $\BB(\Bx)$, and $\BA(\Bx)$ are constant. The simplest 
are ``plane-wave'' solutions where $\Bu(\Bx)=\Bu(\Bk\cdot\Bx)$ and the wave vector $\Bk$, which is not necessarily real, 
has to be chosen appropriately. One may also try using separation of variables in different co-ordinate systems. In
two dimensions there are many exact solutions that involve analytic functions, notably those of 
Muskhelishvili \cite{Muskhelishvili:1963:SBP} and Lekhnitskii \cite{Lekhnitskii:1968:AP}.
Even in three dimensions one can express any solution to Laplace's equation in a convex body in terms of
analytic functions as shown in Section 6.9 of \cite{Milton:2002:TOC}. Green functions with sources in
complex space (i.e. with singularities at points $\Bx=\Bx_0$ where the elements of $\Bx_0$ are complex)
give useful representations of fields that can sometimes be more suitable than standard
representations using, for example, spherical harmonics \cite{Deschamps:1971:GBB,Norris:1997:ECS}.

 Here
we are interested in exact solutions where the coefficients  $\BL(\Bx)$, $\BB(\Bx)$, and $\BA(\Bx)$ are not
constant, which is of course of great practical interest. Many exact solutions can be obtained by mapping over 
the ideas presented in chapters 3, 4, 5, 6, 7, 8, 9, 14 and 17 in the book of Milton \cite{Milton:2002:TOC}
for periodic materials, to materials that are not periodic. These ideas are due to many people, besides myself, so rather than
give a very lengthy list, the reader is referred to the many references in my book. 

A vast array of solutions to the equations \eq{2.1} can be obtained by a change of variable $\By=\Bf(\Bx)$, mapping a solution,
or set of solutions, we know (perhaps in a homogeneous medium) into a solution or set of solutions, with new
coefficients $\BL'(\By)$, $\BB'(\By)$, and $\BA'(\By)$, and a new source $\Br'(\By)$. For electromagnetism
the major realization of Dolin \cite{Dolin:1961:PCT} was that one could interpret this new equation as solving the equations in a new medium with a 
new anisotropic permittivity $\BGve(\By)$ and a new anisotropic permeabilty $\BGm(\By)$: transformation optics, or equivalently transformation
electromagnetism, was born. (An english translation of his paper is available at \url{http://www.math.utah.edu/~milton/DolinTrans2.pdf})
Among other things, Dolin realized that by making
this transformation in empty space one could obtain an inclusion that was invisible to all applied fields. Transformation ideas were also used in \cite{Derrick:1979:CGT, Chandezon:1980:NTM}
to map doubly periodic gratings onto flat gratings. Nicorovici, McPhedran, and Milton
\cite{Nicorovici:1994:ODP} found that annular regions with
a dielectric constant of $-1$ surrounded by a medium of dielectric constant $1$ should be invisible to (almost) all applied quasistatic fields. Al{\'u} and Engheta
\cite{Alu:2005:ATP} found that coated spheres at certain frequencies could be invisible to incident plane waves. Inclusions were found
\cite{Luo:2009:RAP,Xi:2009:ODP} that are invisible to plane waves in one direction and that in fact can cloak objects inside them, as experimentally confirmed by 
Landy and Smith \cite{Landy:2013:FPU}. An earlier paper by Kerker \cite{Kerker:1975:IB} had established that
confocal coated ellipsoids could be invisible to long wavelength waves: these are called neutral inclusions and
there is a large literature concerning them: see, for example, Section 7.11 in  \cite{Milton:2002:TOC},
and more recent citations of these papers.

Transformation
conductivity was discovered by Luc Tartar, who remarked upon it to Kohn and Vogelius \cite{Kohn:1984:IUC}. In section 8.5 of my book \cite{Milton:2002:TOC} I showed
how transformation conductivity could be used
to obtain a vast array of periodic materials for which one could exactly solve the ``cell problem'', i.e., obtain periodic solutions for the current
and electric field that have nonzero averages over the unit cell, and consequently obtain exact formulae for the effective conductivity tensor. 
By making the singular transformation 
\beq \By=\Bf(\Bx) =  \left(\frac{|\Bx|}{2}+1\right)\frac{\Bx}{|\Bx|}\quad{\rm if}~~2>|\Bx|>0,
\eeq{2.20}
that maps a punctured ball of radius $2$ to a hollow shell of outer radius $2$ and inner radius $1$, 
Greenleaf, Lassas, and Uhlmann \cite{Greenleaf:2003:ACC,Greenleaf:2003:NCI} 
realized one could cloak objects: not make them visible to any external probing current fields (see also \cite{Kohn:2008:CCV}). In the transformed geometry conducting objects within the unit
ball $|\By|=1$ feel no probing current field, and therefore are invisible.
This marked the discovery of transformation based cloaking. Transformation ideas also extend to geometric optics 
\cite{Leonhardt:2006:OCM,Leonhardt:2006:NCI,Leonhardt:2006:GRE,Leonhardt:2009:BIN,Leonhardt:2008:FCT}; 
to elasticity \cite{Milton:2006:CEM, Brun:2009:ACI, Norris:2011:ECT, Vasquez:2013:SRP}; to
acoustics \cite{Cummer:2007:PAC,Chen:2007:ACT,Farhat:2008:BCA,Norris:2008:ACT,Greenleaf:2008:ITO,Chen:2010:ACT}; to water waves \cite{Farhat:2008:BCA}; 
to matter waves \cite{Zhang:2008:ISE,Greenleaf:2008:ITO,Greenleaf:2012:CEA}; to plasmonics \cite{Kadic:2010:TPC}; 
to thermal conduction \cite{Jacob:2012:TTC,Schittny:2013:ETT} and to flexural waves 
\cite{Farhat:2009:UEC,Colquitt:2014:TEC,Stenger:2016:EEC}. In some sense the transformation idea, that began with the transformation optics work of Dolin \cite{Dolin:1961:PCT}, 
is a generalization of the idea of using conformal transformations for preserving solutions to the two-dimensional Laplace's equation that is so well known.

Besides the exact transformations for the moduli, approximate transformations for the moduli have been developed too. Often these work quite well:
see, for example, \cite{Cai:2007:OCM,Cai:2007:NMC,Chang:2011:CEW}. Additionally, transformations using complex coordinates have proved useful \cite{Popa:2011:GFH,Castaldi:2013:PTM}.

Curiously, stimulated by a remark of Alexei Efros (private communication, 2005), Nicolae Nicorovici and myself found in 2005 that anomalous resonance creates cloaking where the cloaking device {\it is outside} the cloaking region, although it was not until 2006 that our work was published \cite{Milton:2006:CEA}.
(I think, though am not certain, that we were the first to use the word ``cloaking'' as a scientific term in the published literature, outside computer science, for hiding an object). Anomalous localized resonance was discovered by 
Nicorovici, McPhedran, and myself \cite{Nicorovici:1994:ODP} and as proved in this 1994 paper provides the mechanism for creating ``ghost sources'',
that on one side of which the field converges to a smooth field having a singularity at the point of the ``ghost source'' while on the other
side of which the field has enormous oscillations. The discovery of ghost sources and anomalous resonance is most clear from 
Figure 2 in that paper, reproduced here as \fig{anomlfig},
- see also equation (12) in the 1994 paper
and the text below it. 
See also the rough unpublished draft \cite{Milton:1994:URP}, written prior to May 1996, that emphasizes the observations made in our published 
paper, and see also \cite{Milton:2005:PSQ} where some errors in \cite{Nicorovici:1994:ODP} are corrected.
[Additionally, see the introduction in \cite{Milton:2013:SEH}]. An account of the history of the discovery of anomalous resonance, 
and of ``ghost sources'', that are apparent singularities caused by anomalous resonance, can be found in
\url{http://www.maths.dur.ac.uk/lms/104/talks/1072milt.pdf}. Anomalous resonance provides one of the most striking examples of unusual 
behavior in linear inhomogeneous systems. It has the curious property that the boundaries of the resonant regions (where the field blows up as the 
loss tends to zero) move as the source moves. Whereas normal resonances are associated with poles, anomalous resonance seems to be 
associated with essential singularities \cite{Nicorovici:1993:TPT,Ammari:2013:STN}.

\begin{figure}[!ht]
\includegraphics[width=0.9\textwidth]{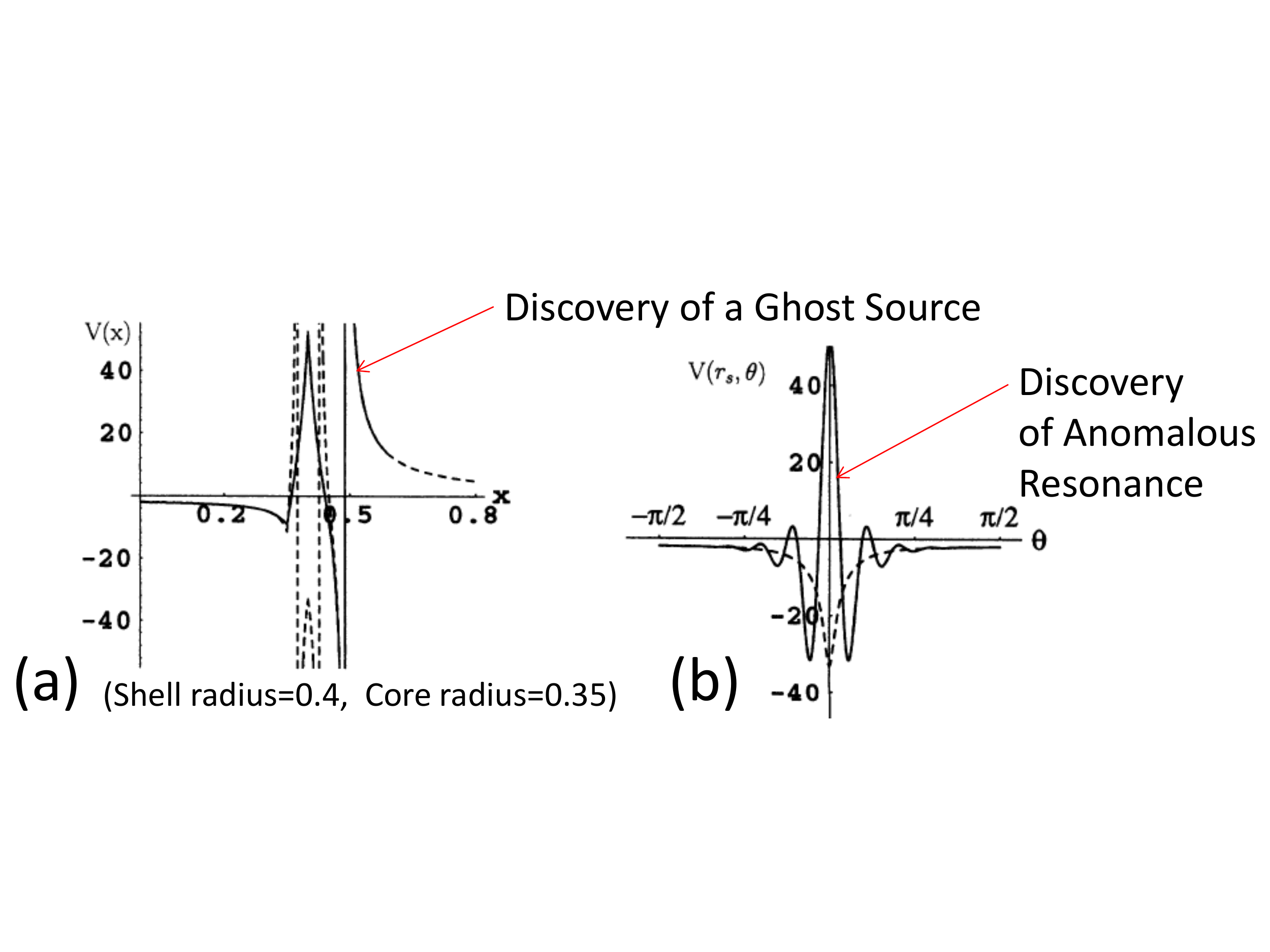}
\vskip -3.0cm
\caption{Reproduction, with permission, of Figure 2 of \protect\cite{Nicorovici:1994:ODP} highlighting our discoveries of ghost sources and anomalous resonance.
The red arrows and accompaning text are new and are inserted to empasize our findings. Note the apparent divergence in the potential at a radius of $0.52$ in Figure (a) 
which is outside the shell radius of 0.40. Also in Figure (b) note the large oscillations of the potential showing the anomalous resonance.}
\labfig{anomlfig}
\end{figure}
Amazingly when the loss is small,
or {\it even when there is no loss}, anomalous resonance has the incredible feature that although energy produced by a suitably placed constant power source when it is turned on propagates in all
directions, at very long times essentially all the energy gets funnelled to the anomalously resonant fields, where it continues to build up with growing fields if there is no loss, 
or is dissipated into heat if there is a small amount of loss \cite{Milton:2006:CEA}--see in particular {\it figures 4 and 5, and the proof in section 3}, also 
see \cite{Milton:2006:OPL,Yaghjian:2006:PWS}). The beautiful subsquent numerical simulations made by Nicolae Nicorovici 
were downloaded so many times that our paper \cite{Nicorovici:2007:QCT}) was in 2007 the most downloaded paper 
in  the journals of the Optical Society of America, with about 13,000 downloads. There is now a substantial body of work that has explored cloaking due to 
anomalous resonance resonance: see \cite{Milton:2008:SFG,Nicorovici:2008:FWC,Nicorovici:2011:RLD,Ammari:2013:ALR}, and the references below.

It is to be emphasized that while the original proofs were for a finite but arbitrary
number of dipole sources, or polarizable particles in two-dimensional quasistatics, or a single dipole source or polarizable particle in three-dimensional quasistatics, or for a single dipole source 
or polarizable particle at finite frequency
in two dimensions \cite{Milton:2006:CEA}), the cloaking extends to particles that are small compared to the wavelength of the anomalous resonance \cite{Bouchitte:2010:CSO},
and to sources of finite size \cite{Ammari:2013:STN, Kohn:2012:VPC, Ammari:2013:STNII, Nguyen:2015:CAL, Meklachi:2016:SAL, Li:2016:QCA}. Sometimes
the cloaking is only partial \cite{Bruno:2007:SCS} or nonexistent \cite{Milton:2006:CEA, Ammari:2013:STN, Kohn:2012:VPC, Onofrei:2016:ALR}; 
and it also extends to more general sources and polarizable particles at finite wavelength \cite{Nicorovici:2008:FWC, Kettunen:2014:AEA, Onofrei:2016:ALR, Nguyen:2016:CVA}.
Most significantly, Nguyen has recently rigorously proved that for two-dimensional quasistatics, arbitrarily shaped 
passive objects of small but finite size that are near the annulus {\it are completely cloaked in the limit as the loss in the system tends to zero}
\cite{Nguyen:2016:EWP}.

Also it is to be emphasized that anomalous resonance and the associated cloaking holds also for magnetoelectric and thermoelectric systems (see section 6 in
\cite{Milton:2005:PSQ}) and for elasticity \cite{Ando:2015:SPN, Li:2016:ALR}, which is why I prefer the term cloaking due
to anomalous resonance rather than plasmonic cloaking \cite{Nicorovici:2008:FWC, McPhedran:2009:CPR}.
While in electromagnetism the anomalous resonance is caused by surface plasmons, surface plasmons do not in general
have the characteristic feature of anomalous resonant fields that (in theory) they are increasingly confined to a localized region dependent of the position of the sources
as the loss in the system goes to zero (or as the time goes to infinity in a loss-less system, with almost time harmonic sources whose amplitude is zero or extremely
close to zero in the far distant past). Cloaking due to complementary media, discovered by Lai, Chen, Zhang, and Chan \cite{Lai:2009:CMI},
is also caused by anomalous resonance: note the highly oscillating fields near the cloaked inclusions
in their numerical simulations. I must admit I was a bit skeptical of this type of cloaking when it first appeared, but Nguyen and
Nguyen \cite{Nguyen:2016:CCM, Nguyen:2015:CAL1} have now given a rigorous proof of this type of cloaking under certain circumstances. 

In 2009 I remembered that the anomalous resonant fields must be caused by polarization charges,
and therefore that one should be able to get the same effect by active sources in a homogeneous medium. This gave birth to the subject of active exterior cloaking
for conductivity \cite{Vasquez:2009:AEC, Vasquez:2012:MAT}, for acoustics in two-dimensions and three-dimensions
\cite{Vasquez:2009:AEC, Vasquez:2009:BEC, Vasquez:2011:ECA, Vasquez:2013:SRP, Norris:2012:SAA}
and for elastodynamics \cite{Norris:2014:AEC}.
In a beautiful twist of the idea, O'Neill, Selsil, McPhedran, Movchan, Movchan, and Moggach \cite{ONeill:2015:ACI, ONeill:2016:ACR} found that for the plate equation one could get 
excellent exterior cloaking without enormous fields if one relaxed the requirement
that the cloak create a quiet zone and instead require only that it cloak a given object. The main reason that enormous fields are not needed is that the
plate equation has the desirable property that the Green's function is bounded.

\section{Metamaterials}
\setcounter{equation}{0}

Metamaterials are basically composite materials, with properties not normally found in nature. They allow a wide range of parameters
$\BL(\Bx)$, $\BB(\Bx)$, and $\BA(\Bx)$ to be realized, thus enabling one to approximately realize many exact solutions. An excellent,
though not completely comprehensive, survey of electromagnetic metamaterials can be found in the slides of Sergei Tretyakov
(\url{https://users.aalto.fi/~sergei/Tretyakov_slides_Metamaterials2015.pdf}). As he mentions,
they have been studied since at least 1898. In particular, Brown \cite{Brown:1953:ADR} realized that media with short wires could give rise
to effective refractive indices less than one. Maz’ya and H{\"a}nler
\cite{Mazya:1993:ASN} realized that one could obtain asymptotic solutions
when part of the domain is cut into tubes that transport the current.
Schelkunoff and Friis \cite{Schelkunoff:1952:ATP} and Lagarkov et.al. \cite{Lagarkov:1997:RPB} realized one could get
artificial magnetism with arrays of split rings (see also \cite{Vinogradov:1997:RPB}). Experiments showing negative magnetic permeabilty
were performed by Lagarkov et.al. \cite{Lagarkov:1997:RPB}. Materials with simultaneous negative permittivity and permeability, and hence a negative
refractive index, that were studied by Veselago \cite{Veselago:1967:ESS}, were experimentally realized by Shelby, Smith, and Schultz \cite{Shelby:2001:EVN}. 
These materials have the curious property that the wave crests move opposite to the group velocity, i.e., opposite to the direction of energy flow.
Waves with such properties were studied as early as 1904 in the works of Lamb \cite{Lamb:1904:GV} and Schuster \cite{Schuster:1904:ITO}.

In fact, in theory, at constant frequency, 
one can get any desired pair $(\BGve,\BGm)$, of permittivity tensor $\BGve$ and permeability tensor $\BGm$ provided they are both 
almost real \cite{Milton:2010:RMP}. Of particular interest are materials where at a given frequency
the dielectric constant is close to zero: these epsilon near zero (ENZ) materials have remarkable
properties \cite{Silveirinha:2006:TEE} and can be used to design new types of theoretical "electromagnetic circuits" \cite{Milton:2009:EC, Milton:2010:HEC}
that operate at a single frequency.
By laminating a material having a negative dielectric constant over a range of frequencies, with a material having a positive dielectric constant one can get
hyperbolic materials where, over a range of frequencies, the dielectric tensor has both positive and negative eigenvalues, and these can be used as hyperlenses 
\cite{Jacob:2006:OHF,Salandrino:2006:FFS,Rho:2010:SHT,Lu:2012:HMF}. In the infrared and optical regime the building blocks for constructing metamaterials have been plasmonic materials
like silver, gold, and silicon carbide that have a negative real part to the permittivity, and a comparatively small imaginary part, though at best it is still
about one tenth of the magnitude of the real part. Now, new better performing plasmonic materials are available \cite{Naik:2013:APM}.

Interestingly, in three dimensions, but not in two-dimensions, one can combine materials with positive Hall-coefficient to obtain a composite with negative Hall-coefficient
\cite{Briane:2009:HTD,Kadic:2015:HES}: see also \cite{Briane:2004:CSC}, thereby destroying the argument that in classical physics it is the sign of the Hall coefficient 
which tells one the sign of the charge carrier. A fantastic experimental confirmation of this result has been obtained in \cite{Kern:2016:EES}.

My own interest in metamaterials was sparked by some brilliant experiments of Roderic Lakes \cite{Lakes:1987:FSN},
that showed it was possible to get isotropic foams with a negative Poisson's ratio that widen as they stretched. Subsequently I obtained the first
rigorous proof \cite{Milton:1992:CMP} that negative Poisson ratio (Auxetic) materials exist within the framework of continuum elasticity, without voids or sliding surfaces.
Although I did not draw attention to it at the time, one model that I came up with, having hexagonal spoked wheels linked by deformable parallelogram
structures, had the interesting property that when the spoken wheels and linkages were almost rigid then its only easy mode of deformation is a dilation
over some (not infinitesimal) window of strains. Another astoundingly simple two-dimensional dilational material consisting of rotating squares,
was discovered by Grima and Evans \cite{Grima:2000:ABR}. It was related to some models that Sigmund \cite{Sigmund:1995:TMP} had found earlier, both in two and three dimensions, 
but which had sliding surfaces. In fact, the rotating square model was known to Pauling \cite{Pauling:1930:SSS},
although he does not appear to have made a connection with Auxeticity.

Three dimensional dilational materials
without sliding surfaces have been discovered more recently \cite{Milton:2013:CCMa,Milton:2015:NET}, 
and approximations of such structures have been built and tested \cite{Buckmann:2014:TDD}. Amazingly, 
any fourth-order tensor that is positive definite and has the symmetries of elasticity tensors, can be realized 
as the effective elasticity tensor of a mixture of a sufficiently compliant phase and a sufficiently stiff phase
\cite{Milton:1995:WET}. The building blocks for the construction are pentamode
materials: elastic materials that essentially can only support one stress, but that stress can be any given matrix. In an incredible feat of three-dimensional
lithography, these have actually been built \cite{Kadic:2012:PPM} and used to construct an ``unfeelability cloak'' \cite{Buckmann:2014:EMU}. 
More generally it is theoretically possible to design materials
whose macroscopically affine deformations are constrained to lie on any desired trajectory in the 6-dimensional deformation space, excluding deformations where
the structure collapses to a lower dimension \cite{Milton:2013:CCMa}). Also a lot of progress has been made on the characterization 
of the possible effective elasticity tensors of three-dimensional
printed materials when the elastic moduli, and volume fraction, of the constituent material are known, assuming there is just one constituent material plus void
\cite{Milton:2016:PEE}.

It is also possible to combine three materials (or two plus void)
all of which by themselves expand when heated to obtain a composite which contracts
when heated, or alternatively which expands more than the constituent materials \cite{Lakes:1996:CSS,Sigmund:1996:CET,Sigmund:1997:DME}. 

One can combine materials with positive mass density
to obtain composites with negative effective mass density
\cite{Sheng:2003:LRS,Liu:2005:AMP}; and one can obtain materials
with, at a given frequency, anisotropic and even complex effective mass density
\cite{Schoenberg:1983:PPS,Milton:2006:CEM,Buckmann:2015:EMU}. In fact, it follows directly from the work of Movchan and Guenneau \cite{Movchan:2004:SRR}
that there is a close link between negative effective mass density in antiplane elasticity and negative magnetic permeability: 
in cylindrical geometries the Helmholtz equation is common to both physical phenomena. Moreover, as pointed out by Milton and Willis
\cite{Milton:2006:MNS}, in metamaterials Newton's 
law $F=ma$ should be replaced by a law where the force depends on the acceleration at at previous times, and not just at the present time 
since it takes some time for the constituents to move: they do not necessarily move together in lock step motion.  

However, with respect to the results in the previous paragraph, attention should be drawn to the pioneering work of Auriault and Bonnet
\cite{Auriault:1985:DCE,Auriault:1994:AHM} who studied homogenization in high contrast elastic media and found that the effective density could be anisotropic and
frequency dependent. To quote them "The monochromatic macroscopic behavior is elastic, but
with an effective density $\Gvr^{\rm eff}$ of tensorial character and depending on the pulsation". Moreover they realized that the 
effective density could be negative. Referring to a figure they say "hatched areas
correspond to negative densities $\Gvr^{\rm eff}$, i.e., to stopping bands." 
A more rigorous approach to the same problem was developed by Zhikov \cite{Zhikov:2004:GSE}. Smyshlyaev \cite{Smyshlyaev:2009:PLe} extended
the analysis to elastic waves propagating in extremely anisotropic media. The extension to
Maxwell's equations was done by Bouchitte and Schweizer
\cite{Bouchitte:2010:HME} and Cherednichenko and Cooper \cite{Cherednichenko:2015:HSH}. 

Noteworthy is that in 1982, McPhedran, Botten, Craig, Nevi{\`e}re and Maystre \cite{McPhedran:1982:LLG} had realized that conventional homogenization was 
not appropriate for high contrast lossy lamellar gratings, even in the long wavelength limit. 
Also, in 1985, in his ensemble average framework for studying waves in composites,
Willis \cite{Willis:1985:NID} proved that his ``effective density'' operator is a second order tensor.

What is especially remarkable is that Camar-Eddine and Seppecher \cite{Camar:2002:CSD,Camar:2003:DCS}) have been able to completely characterize, 
in a certain sense, all possible homogenized behaviors, including non-local ones, in mixtures of high
contrast conducting materials and high contrast linear elastic materials. The key elements are dumbbell shaped inclusions, on many scales. The highly conducting dumbbells,
or almost rigid dumbbells, provide a non-local interaction between the balls at the ends when the radius of the bar joining the balls shrinks to zero. 
The material near the bar is affected by the potential or displacement
of the bar, but this scales away as the radius of the bar shrinks to zero. So one can treat it as if the bar is absent, but with a non-local interaction 
between the balls.

At the discrete level, all possible dynamic behaviors of linear mass-spring networks have been
completely characterized \cite{Vasquez:2011:CCS}. This gives hope that a simllar characterization may be possible in the continuum case. Whether such a characterization will have
any practical  significance is not at all clear. In the case of the paper of Camar-Eddine and Seppecher \cite{Camar:2003:DCS} at any finite deformation non-linear effects are important in the delicate structures that are needed to achieve some of the exotic linear behaviors.

Specific examples
of exotic behaviors in high contrast elastic systems have been given by Seppecher, Alibert, and dell'Isola \cite{Seppecher:2011:LET}. Other exotic behaviors of high contrast composites
have been found and explored \cite{Khruslov:1978:ABS,Briane:1998:HSW,Briane:1998:HTR,Belov:2003:SSD,Cherednichenko:2006:NLH,Shin:2007:TDE,Milton:2007:NMM,Kaelberer:2010:TDR}.
The architect/artist Boris Stuchebryukov has constructed incredible metamaterials made from parallelogram mechanisms using razor blades (see the beautiful videos \url{https://www.youtube.com/watch?v=l8wrT2YB5s8} and 
\url{https://www.youtube.com/watch?v=AkUn8nFd8mk}).

We mention too, that part of the reason for the surge of interest in these metamaterials arises because precision
three-dimensional lithography and printing techniques \cite{Kadic:2012:PPM,Buckmann:2012:TMM,Buckmann:2014:TDD,Meza:2014:SLR} now allow one to
tailor beautiful structures with desired properties. Still there are limitations: usually one wants the cell size to be
small and this restricts the size of the overall sample, since in three dimensions the number of cells scales as the cube 
of the sample side. For this reason metasurfaces and planar lattice materials may hold more promise for practical applications.

\section{Analytic Materials of Type One}
\setcounter{equation}{0}
Suppose ${\bf x}\in \mathbb{R}^d$ and 
one has a basis of $d$-complex wavevectors $\Bk^{(\ell)}$, $\ell=1,2,\ldots,d$. Suppose further that
\beq L_{j\Ga i \Gb}(\Bx)k^{(\ell)}_i=P^{(\ell)}_{\Ga j \Gb}(\Bx\cdot\Bk^{(\ell)}),\quad
B_{\Ga i \Gb}(\Bx)k^{(\ell)}_i=Q^{(\ell)}_{\Ga\Gb}(\Bx\cdot\Bk^{(\ell)}), \quad
A_{\Ga\Gb}(\Bx)=A_{\Ga\Gb},\quad r_{\Ga}(\Bx)=\sum_{\ell}S_{\Ga\ell}(\Bx\cdot\Bk^{(\ell)}),
\eeq{3.1}  
where the coefficients $A_{\Ga\Gb}$ are independent of $\Bx$, and $P^{(\ell)}_{i\Ga\Gb}(z)$, $Q^{(\ell)}_{\Ga\Gb}(z)$
and $S_{\Gb\ell}(z)$ are analytic functions of $z$. Note that since the $\Bk^{(\ell)}$ are a basis, \eq{3.1}
completely determines all elements of $\BL(\Bx)$ and $\BB(\Bx)$. 

We look for a solution $\Bu(\Bx)$ with elements $u_\Gb$  of the form
\beq u_\Gb(\Bx)=\sum_{\ell}f_{\Gb\ell}(\Bx\cdot\Bk^{(\ell)}), \eeq{3.2}
where the $f_{\Gb\ell}(z)$ are analytic functions of $z$. Then we have 
\beq \frac{\Md u_\Gb(\Bx)}{\Md x_i}=\sum_{\ell}k_i^{(\ell)}f'_{\Gb\ell}(\Bx\cdot\Bk^{(\ell)}), \eeq{3.3}
and consequently
\beq  L_{j\Ga i \Gb}(\Bx) \frac{\Md u_\Gb(\Bx)}{\Md x_i}=
\sum_\ell L_{j\Ga i \Gb}(\Bx)k_i^{(\ell)}f'_{\Gb\ell}(\Bx\cdot\Bk^{(\ell)})=
\sum_\ell P^{(\ell)}_{\Ga j\Gb}(\Bx\cdot\Bk^{(\ell)})f'_{\Gb\ell}(\Bx\cdot\Bk^{(\ell)}), \eeq{3.4}
which implies 
\beq \frac{\Md}{\Md x_j}\left[L_{j\Ga i \Gb}(\Bx)\frac{\Md u_\Gb(\Bx)}{\Md x_i}\right]
=\sum_{\ell}k_j^{(\ell)}\left[P^{(\ell)}_{\Ga j\Gb}(z)f'_{\Gb\ell}(z)\right]'\biggr\rvert_{z=\Bx\cdot\Bk^{(\ell)}}.
\eeq{3.5}
We also have
\beq B_{\Ga i \Gb}(\Bx)\frac{\Md u_\Gb(\Bx)}{\Md x_i}=
\sum_\ell Q^{(\ell)}_{\Ga\Gb}(\Bx\cdot\Bk^{(\ell)})f'_{\Gb\ell}(\Bx\cdot\Bk^{(\ell)}). \eeq{3.6}
By substituting these expressions back in our main equation \eq{2.2} we arrive at
\beq \sum_{\ell}k_j^{(\ell)}\left[P^{(\ell)}_{\Ga j\Gb}(z)f'_{\Gb\ell}(z)\right]'+ Q^{(\ell)}_{\Ga\Gb}(z)f'_{\Gb\ell}(z)
+A_{\Ga\Gb}f_{\Gb\ell}(z)-S_{\Ga\ell}(z)\biggr\rvert_{z=\Bx\cdot\Bk^{(\ell)}}=0.
\eeq{3.7}
So the $f_{\Gb\ell}(z)$ should satisfy
\beq k_j^{(\ell)}\left[P^{(\ell)}_{\Ga j\Gb}(z)\right]'f'_{\Gb\ell}(z)
+k_j^{(\ell)}P^{(\ell)}_{\Ga j\Gb}(z)f''_{\Gb\ell}(z)
+ Q^{(\ell)}_{\Ga\Gb}f'_{\Gb\ell}(z)
+A_{\Ga\Gb}f_{\Gb\ell}(z)-S_{\Ga\ell}(z)=0,
\eeq{3.8}
for $\Ga=1,2,\ldots,m$, where sums over the repeated indices $j$ and $\Gb$ are implied. For fixed $\ell$ this
provides $m$ ordinary differential equations for the $m$ unknowns $f_{\Gb\ell}(z)$, $\Gb=1,2,\ldots,m$.
We can rewrite \eq{3.8} in matrix form as
\beq \left[\Bk^{(\ell)}\cdot\BP^{(\ell)}(z)\right]'\left[\Bf^{(\ell)}(z)\right]'
+\Bk^{(\ell)}\cdot\BP^{(\ell)}(z)\left[\Bf^{(\ell)}(z)\right]''+\BQ^{(\ell)}(z)\left[\Bf^{(\ell)}(z)\right]'+\BA\Bf^{(\ell)}(z)
=\Bs^{(\ell)}(z),
\eeq{3.8aa}
where for a given $\ell$, $\Bf^{(\ell)}(z)$ and $\Bs^{(\ell)}(z)$ are at each point $z$, $m$-component vectors with elements $f_{\Gb\ell}(z)$ and $S_{\Gb\ell}(z)$, $\Gb=1,2,\ldots,m$; $\BQ^{(\ell)}(z)$ and $\BA$ 
are $m\times m$ matrices with elements $Q^{(\ell)}_{\Ga\Gb}(z)$ and $A_{\Ga\Gb}$; 
$\BP^{(\ell)}(z)$ is a 3-index object with one space index and two field indices-- the $d$-dimensional vector
$\Bk^{(\ell)}$ acts on the space index of $\BP^{(\ell)}(z)$ leaving a $m\times m$ matrix $\Bk^{(\ell)}\cdot\BP^{(\ell)}(z)$ with elements
$k_j^{(\ell)}P^{(\ell)}_{\Ga j\Gb}(z)$.

Note that it could be the case that \eq{3.1}, or equivalently \eq{3.8aa} only holds for $\ell=1,2,\ldots, p$ where $p<d$. In that case we call the material an ``incomplete analytic material  of type 1, and rank $p$. In the case where $p=d$ we have a 
``complete analytic material of type 1'', or just simply, an ``analytic material of type 1''.

Analytic materials are easily generated from \eq{3.8aa}. The simplest way to do this is just to arbitrarily pick $d^2m^2$
analytic functions $P^{(\ell)}_{\Ga j\Gb}(z)$, $dm^2$ analytic functions $Q^{(\ell)}_{\Ga\Gb}(z)$, $m^2$ constants
$A_{\Ga\Gb}$ and $md$ analytic functions $f_{\Gb\ell}(z)$, taking care to ensure that the resultant 
singularities in spatial coordinates lie inside the body $\GO$. Then choosing the $d^2$ constants $k_j^{(\ell)}$
so that the vectors $\Bk^{(\ell)}$, $\ell=1,2,\ldots,d$ form a basis, we can ensure that \eq{3.8aa} is satisfied by
choosing the left hand side to be our source $\Bs^{(\ell)}(z)$ that is then analytic in $z$. Moreover if the given
analytic functions are polynomials, then the matrix elements $S_{\Ga\ell}(z)$ will also be polynomials. Such a material
would then be a polynomial material -- defined as one for which the coefficients and sources are entirely expressed in terms of polynomials, and are such that there is at least one solution for $\Bu(\Bx)$ involving polynomials $f_{\Gb\ell}(z)$.

Even when there are no sources, i.e., when $\Bs^{(\ell)}(z)=0$ for all $\ell$, we could, for example,
ensure that \eq{3.8aa} is satisfied by picking a set of $dm$ analytic functions $e_{\Gg\ell}(z)$, $\ell=1,\ldots,d$, $\Gg=1,\ldots,m$ and
choosing the matrix elements $Q^{(\ell)}_{\Ga\Gb}(z)$ to be given by
\beq Q^{(\ell)}_{\Ga\Gb}=-\frac{e_{\Gb\ell}(z)}{e_{\Gg\ell}(z)f'_{\Gg\ell}(z)}
\left\{k_j^{(\ell)}[P^{(\ell)}_{\Ga j\Gn}(z)]'f'_{\Gn\ell}(z)+k_j^{(\ell)}P^{(\ell)}_{\Ga j\Gn}(z)f''_{\Gn\ell}(z)+A_{\Ga\Gn}f_{\Gn\ell}(z)\right\},
\eeq{3.8ab}
where summation over the repeated indices $\Gg$, $j$, and $\Gn$ is implied, but no sum is taken over $\ell$.

Undoubtedly, there are many more ways of generating analytic materials. The preceding examples just serve to demonstrate that there exist a wide class of
analytic materials. 

\section{Analytic Materials of Type Two}
\setcounter{equation}{0}
Suppose one has a basis $\Be^{\ell}$, $\ell=1,2,\ldots,md$ of $d\times m$ matrices such that each
$\Be^{\ell}$ is a rank one matrix:
\beq e^{\ell}_{i\Gb}=k^{(\ell)}_iv^{(\ell)}_\Gb,
\eeq{3.9}
where the $k^{(\ell)}_i$ and $v^{(\ell)}_\Gb$ are complex and do not depend on $\Bx$. Assume further that
\beqa L_{j\Ga i \Gb}(\Bx)e^{(\ell)}_{i\Gb} & = & P^{(\ell)}_{j\Ga}(\Bx\cdot\Bk^{(\ell)}),\quad
B_{\Ga i \Gb}(\Bx)e^{(\ell)}_{i\Gb}=Q^{(\ell)}_{\Ga}(\Bx\cdot\Bk^{(\ell)}), \nonum
A_{\Ga\Gb}(\Bx)v^{(\ell)}_\Gb & = & H_{\Ga}^{(\ell)}(\Bx\cdot\Bk^{(\ell)}),\quad r_{\Ga}(\Bx)=\sum_{\ell}S_{\Ga}^{(\ell)}(\Bx\cdot\Bk^{(\ell)}),
\eeqa{3.10}  
where the $P^{(\ell)}_{j \Ga}(z)$, $Q^{(\ell)}_{\Ga}(z)$, $H_{\Ga}^{(\ell)}(z)$ and $S_{\Ga\ell}(z)$ are analytic functions of $z$
but should not be confused with the functions in the previous section. 

Now we look for a solution
\beq u_\Gb=\sum_{\ell}v^{(\ell)}_\Gb f_\ell(\Bx\cdot\Bk^{(\ell)}). \eeq{3.11}
We get 
\beq \sum_{\ell}k_j^{(\ell)}\left[P^{(\ell)}_{j \Ga}(z)f_\ell'(z)\right]'+ Q^{(\ell)}_{\Ga}(z)f_\ell'(z)
+H^{(\ell)}_{\Ga}(z)f_\ell(z)-S_{\Ga}^{(\ell)}(z)\biggr\rvert_{z=\Bx\cdot\Bk^{(\ell)}}=0.
\eeq{3.12}
So each function $f_{\ell}(z)$ should solve the $m$ second-order differential equations, 
\beq k_j^{(\ell)}\left[P^{(\ell)}_{\Ga j}(z)\right]'f'_{\ell}(z)
+k_j^{(\ell)}P^{(\ell)}_{\Ga j}(z)f''_{\ell}(z)+ Q^{(\ell)}_{\Ga}(z)f'_{\ell}(z)
+H^{(\ell)}_{\Ga}(z)f_{\ell}(z)=S_{\Ga}^{(\ell)}(z), \eeq{3.13}
for $\Ga=1,2,\ldots,m$. This is generally an overdetermined system of equations, as we have $m$ equations to be satisfied
by a single function $f_{\ell}(z)$. So let us make the additional assumption that there exist real or complex
constants $c_\Ga$ and analytic functions $p^{(\ell)}_{j}(z)$, $q^{(\ell)}(z)$, $h^{(\ell)}(z)$, and  $s^{(\ell)}(z)$ such that
\beq P^{(\ell)}_{\Ga j}(z)=c_\Ga p^{(\ell)}_{j}(z),\quad
Q^{(\ell)}_{\Ga}(z)=c_\Ga q^{(\ell)}(z),\quad H^{(\ell)}_{\Ga}=c_\Ga h^{(\ell)}(z), \quad S_{\Ga}^{(\ell)}(z)=c_\Ga s^{(\ell)}(z).
\eeq{3.14}
Then the fundamental equations \eq{3.13} reduce to 
\beq  k_j^{(\ell)}\left[p^{(\ell)}_{j}(z)\right]'f'_{\ell}(z)
+k_j^{(\ell)}p^{(\ell)}_{j}(z)f''_{\ell}(z)+ q^{(\ell)}(z)f'_{\ell}(z)
+h^{(\ell)}(z)f_{\ell}(z)=s^{(\ell)}(z).
\eeq{3.15}

Again it could be the case that \eq{3.10} and \eq{3.14} only holds for $\ell=1,2,\ldots, p$ where $p<md$. In that case
we call the material an ``incomplete analytic material of type 2, and rank $p$. In the case where $p=md$ we have a 
``complete analytic material of type 2'', or just simply, an ``analytic material of type 2''.

As for analytic materials of type one, solutions to \eq{3.15} are easily generated. For example, one could just simply arbitrarily choose
all  the analytic functions aside from those functions $s^{(\ell)}(z)$ determining the sources (making sure the resultant singularities in spatial coordinates lie the body $\GO$) and then
let the left hand side of \eq{3.15} determine these sources. In the absence of sources one could arbitrarily choose the analytic functions $p^{(\ell)}_{j}(z)$, $q^{(\ell)}(z)$, and $f_{\ell}(z)$ and then
to satisfy \eq{3.15} choose 
\beq h^{(\ell)}(z)=-\left\{ k_j^{(\ell)}[p^{(\ell)}_{j}(z)]'f'_{\ell}(z)
+k_j^{(\ell)}p^{(\ell)}_{j}(z)f''_{\ell}(z)+ q^{(\ell)}(z)f'_{\ell}(z)\right\}  /f_{\ell}(z),
\eeq{3.15a}
where the sum over $j$ is implied, but not over $\ell$. Alternatively, if $h^{(\ell)}(z)$ is given, but not $q^{(\ell)}(z)$, then we can divide \eq{3.15}
by $f'_{\ell}(z)$ to obtain a formula for $q^{(\ell)}(z)$ in terms of the other analytic functions.

\section{Analytic Materials of Type Three}
\setcounter{equation}{0}

Let us again begin with the equations 
\beq 
\frac{\Md}{\Md x_j}\left[L_{j\Ga i \Gb}(\Bx)\frac{\Md u_\Gb(\Bx)}{\Md x_i}\right]
+B_{\Ga i \Gb}(\Bx)\frac{\Md u_\Gb(\Bx)}{\Md x_i}
+A_{\Ga\Gb}(\Bx)u_\Gb(\Bx)=r_{\Gb}(\Bx), \eeq{4.1}
and look for solutions of the form
\beq u_\Gb(\Bx)=f_\Gb(t(\Bx)), \eeq{4.2}
where $f_\Gb(z)$ is an analytic function of $z$, and $t(\Bz)$ 
is some twice differentiable function mapping $\GO$ onto a region $\GU$ of the complex plane.
We have
\beqa \frac{\Md u_\Gb(\Bx)}{\Md x_i}
& = & \frac{\Md t(\Bx)}{\Md x_i}f'_\Gb(t(\Bx)), \nonum
\frac{\Md^2 u_\Gb(\Bx)}{\Md x_i\Md x_j}
& = &  
\frac{\Md t(\Bx)}{\Md x_i} \frac{\Md t(\Bx)}{\Md x_j}
f''_\Gb(t(\Bx)+
\frac{\Md^2 t(\Bx)}{\Md x_i\Md x_j}f'_\Gb(t(\Bx)).
\eeqa{4.3}
Then by simple differentiation,
\beqa &~& \frac{\Md}{\Md x_j}\left[L_{j\Ga i \Gb}(\Bx)\frac{\Md u_\Gb(\Bx)}{\Md x_i}\right]
 =  \left[\frac{\Md}{\Md x_j}L_{j\Ga i \Gb}(\Bx)\right]\frac{\Md u_\Gb(\Bx)}{\Md x_i}
+L_{j\Ga i \Gb}(\Bx)\frac{\Md^2 u_\Gb(\Bx)}{\Md x_i\Md x_j} \nonum
&~& =  \left[\frac{\Md}{\Md x_j}L_{j\Ga i \Gb}(\Bx)\right]\frac{\Md t(\Bx)}{\Md x_i}f'_\Gb(t(\Bx))
+L_{j\Ga i \Gb}(\Bx)\frac{\Md t(\Bx)}{\Md x_i} \frac{\Md t(\Bx)}{\Md x_j}
f''_\Gb(t(\Bx)
+L_{j\Ga i \Gb}(\Bx)\frac{\Md^2 t(\Bx)}{\Md x_i\Md x_j}f'_\Gb(t(\Bx)),\nonum
&~&
\eeqa{4.4}
and 
\beq B_{\Ga i \Gb}(\Bx)\frac{\Md u_\Gb(\Bx)}{\Md x_i}
=B_{\Ga i \Gb}(\Bx)\frac{\Md t(\Bx)}{\Md x_i}f'_\Gb(t(\Bx)),
\eeq{4.5}
Introducing a set of $m$ functions $h_\Ga(\Bx)$ that multiply the equations, the equations that interest us now take the form,
\beqa &~&\left[h_\Ga(\Bx)L_{j\Ga i \Gb}(\Bx)\frac{\Md t(\Bx)}{\Md x_i} \frac{\Md t(\Bx)}{\Md x_j}\right]f''_\Gb(t(\Bx))\nonum
&~&+\left[h_\Ga(\Bx)\frac{\Md}{\Md x_j}L_{j\Ga i \Gb}(\Bx)\frac{\Md t(\Bx)}{\Md x_i}
+h_\Ga(\Bx)L_{j\Ga i \Gb}(\Bx)\frac{\Md^2 t(\Bx)}{\Md x_i\Md x_j}
+h_\Ga(\Bx)B_{\Ga i \Gb}(\Bx)\frac{\Md t(\Bx)}{\Md x_i}\right]f'_\Gb(t(\Bx))\nonum
&~&+h_\Ga(\Bx)A_{\Ga\Gb}(\Bx)f_\Gb(t(\Bx))
=h_\Ga(\Bx)r_\Ga(\Bx),
\eeqa{4.6}
for $\Ga=1,2,\ldots,m$, where there is no sum over $\Ga$, even though it is a repeated index. Now suppose the
coefficients in the equations are such that for some analytic functions 
$G^L_{\Ga\Gb}(z)$, $G^L_{\Ga\Gb}(z)$, $G^B_{\Ga\Gb}(z)$, $G^A_{\Ga\Gb}(z)$, $g^r_{\Ga}(z)$ we have
\beqa \left[h_\Ga(\Bx)L_{j\Ga i \Gb}(\Bx)\frac{\Md t(\Bx)}{\Md x_i} \frac{\Md t(\Bx)}{\Md x_j}\right]
& = & G^L_{\Ga\Gb}(t(\Bx)),\nonum
\left[h_\Ga(\Bx)\frac{\Md}{\Md x_j}L_{j\Ga i \Gb}(\Bx)\frac{\Md t(\Bx)}{\Md x_i}
+h_\Ga(\Bx)L_{j\Ga i \Gb}(\Bx)\frac{\Md^2 t(\Bx)}{\Md x_i\Md x_j}
+h_\Ga(\Bx)B_{\Ga i \Gb}(\Bx)\frac{\Md t(\Bx)}{\Md x_i}\right]& = & G^B_{\Ga\Gb}(t(\Bx)),\nonum
h_\Ga(\Bx)A_{\Ga\Gb}(\Bx)& = & G^A_{\Ga\Gb}(t(\Bx)),\nonum
h_\Ga(\Bx)r_\Ga(\Bx)& = & g^r_{\Ga}(t_\Ga(\Bx)),
\eeqa{4.7}
where again there is no sum over the repeated index $\Ga$. 
Then the equations \eq{4.6} reduce to
\beq G^L_{\Ga\Gb}(z)f''_\Gb(z)+G^B_{\Ga\Gb}(z)f'_\Gb(z)+G^A_{\Ga\Gb}(z)f_\Gb(z)=g^r_{\Ga}(z), \eeq{4.8}
with $z=t(\Bx)$. We can rewrite \eq{4.8} in the more compact form
\beq \BG^L(z)\Bf''(z)+\BG^B(z)\Bf'(z)+\BG^A(z)\Bf(z)=\Bg^r(z), \eeq{4.9}
where $\BG^L(z)$, $\BG^B(z)$, and $\BG^A(z)$ are $m\times m$ matrix-valued analytic
functions of $z$, by which we mean that their elements $G^L_{\Ga\Gb}(z)$,
$G^B_{\Ga\Gb}(z)$, and $G^A_{\Ga\Gb}(z)$ are analytic functions of $z$, and
$\Bg^r(\Bz)$ and $\Bf(z)$ are $m$-dimensional vector valued analytic
functions of $z$, by which we mean that their elements $g^r_\Ga(\Bz)$ and $f_\Ga(z)$
are analytic functions of $z$. If we have solutions to the system of ordinary differential equations
\eq{4.8}, say for real values of $z$,  and if these solutions can be extended to the region $\GU$ of the complex-plane
without any singularities in  $\GU$, then automatically by substituting
$\Bz=t(\Bx)$ we have a solututions to the original inhomogeneous partial differential equation.

Once again, analytic functions that satisfy \eq{4.8} are easily generated: sources could be chosen with $g^r_{\Ga}(z)$
determined by the left hand side of \eq{4.8}. Or, if the sources are zero, one could pick $m$ analytic functions $e_{\Gb}(z)$,
$\Gb=1,2,\ldots,m$, and set 
\beq G^A_{\Ga\Gb}(z)=-\frac{e_{\Gb}(z)}{e_{\Gg}(z)f_{\Gg}(z)}
\left[G^L_{\Ga\Gn}(z)f''_\Gn(z)+G^B_{\Ga\Gn}(z)f'_\Gn(z)\right],
\eeq{4.10}
in which sums over $\Gg$ and $\Gn$ are implied, and the $f_{\Ga}(z)$, $G^L_{\Ga\Gn}(z)$, and $G^B_{\Ga\Gn}(z)$ are given
analytic functions. Having determined these analytic functions there is a lot of flexibility in choosing coefficients
$L_{j\Ga i \Gb}(\Bx)$, $B_{\Ga i \Gb}$ and $A_{\Ga\Gb}(\Bx)$ so that \eq{4.7} is satisfied.

\section{Analytic Materials in Two-Dimensions}
\setcounter{equation}{0}
Two-dimensions is rather special in that if for some vector field $\Bj(\Bx)$, $\Div\Bj(\Bx)=0$ in a region $\GO$, then we have 
$\Curl\Be=0$, where 
\beq \Be=\BR^\perp\Bj,\quad{\rm and}\quad \BR^\perp=\bpm 0 & 1 \\ -1 & 0 \epm \eeq{5.1}
is the matrix for a $90^\circ$ pointwise rotation of the vector fields, and in two-dimensions $\Curl\Be$ is defined to be
scalar field 
\beq \Curl\Be\equiv\frac{\Md e_2}{\Md x_1}-\frac{\Md e_1}{\Md x_2}. \eeq{5.2}
This is rather easy to see: it follows directly from \eq{5.1} that $e_1(\Bx)=j_2(\Bx)$, and $e_2(\Bx)=-j_1(\Bx)$, and by
substition in \eq{5.2} we see that $\Div\Bj(\Bx)=0$ implies $\Curl\Be=0$. If the region $\GO$ is simply connected it 
then follows that $\Be=-\Grad w$, or equivalently that $\Bj=\BR^\perp\Grad w$. 

So consider the equation 
\beq \Div\BL(\Bx)\Grad\BV(\Bx)+\BB(\Bx)\cdot\Grad\BV(\Bx)
+\BA(\Bx)\BV(\Bx)=\Br(\Bx), \eeq{5.3}
where $\BV(\Bx)$ is an $m$-component potential. Suppose further that $\BB(\Bx)$, $\BA(\Bx)$, and $\Br(\Bx)$ are such that
\beq \BB(\Bx)\cdot\Grad\BV(\Bx)-\BA(\Bx)\BV(\Bx)-\Br(\Bx)=\Div[\BD(\Bx)\BV(\Bx)-\Bt(\Bx)],
\eeq{5.4}
or in components,
\beqa B_{\Ga i \Gb}(\Bx)\frac{\Md V_\Gb(\Bx)}{\Md x_i}
+A_{\Ga\Gb}(\Bx)V_\Gb(\Bx)-r_{\Gb}(\Bx) 
& = & \frac{\Md}{\Md x_i}[D_{i \Ga \Gb}(\Bx)V_{\Gb}(\Bx)-t_{i\Ga}(\Bx)] \nonum
& = & D_{i \Ga \Gb}(\Bx)\frac{\Md V_\Gb(\Bx)}{\Md x_i}
\left[\frac{\Md}{\Md x_i}D_{i \Ga \Gb}(\Bx)\right]V_{\Gb}(\Bx)
-\frac{\Md t_{i \Ga}(\Bx)}{\Md x_i}.\nonum
&~&
\eeqa{5.5}
Clearly this is satisfied when
\beq B_{\Ga i \Gb}(\Bx)=D_{i \Ga \Gb}(\Bx),\quad A_{\Ga\Gb}(\Bx)=\frac{\Md}{\Md x_i}D_{i \Ga \Gb}(\Bx),\quad
r_{\Gb}(\Bx)=\frac{\Md t_{i \Ga}(\Bx)}{\Md x_i}.
\eeq{5.6}
Of course the coefficients $\BD(\Bx)$ and $\Bt(\Bx)$ are not uniquely determined given $\BB(\Bx)$, $\BA(\Bx)$, and
$\Br(\Bx)$ such that \eq{5.6} holds. The easiest way to ensure that \eq{5.6} holds is of course to choose 
$\BD(\Bx)$ and $\Bt(\Bx)$ and then let \eq{5.6} determine $\BB(\Bx)$, $\BA(\Bx)$, and $\Br(\Bx)$. We are then left
with the equation
\beq \Div\left[\BL(\Bx)\Grad\BV(\Bx)+\BD(\Bx)\BV(\Bx)-\Bt(\Bx)\right]=0, \eeq{5.7}
that if $\GO$ is simply connected can be rewritten as
\beq \BL(\Bx)\Grad\BV(\Bx)+\BD(\Bx)\BV(\Bx)-\Bt(\Bx)=\BR^\perp\Grad\BW(\Bx), \eeq{5.8}
where $\BW(\Bx)$ is an $m$-component vector potential. Multiplying this by $\BE(\Bx)$ where $\BE(\Bx)$ is a fourth
order tensor with elements $E_{k \Gg j\Ga}$, $k,j=1,\ldots,d$, $\Gg,\Ga=1,\ldots, m$ gives
\beq E_{k \Gg j\Ga}(\Bx)L_{j\Ga i \Gb}(\Bx)\frac{\Md V_\Gb(\Bx)}{\Md x_i}+E_{k \Gg j\Ga}(\Bx)D_{j\Ga\Gb}(\Bx)V_{\Gb}(\Bx)-E_{k \Gg i\Ga}(\Bx)t_{i \Ga}(\Bx)
=E_{k \Gg j\Ga}R^\perp_{ji}\frac{\Md W_\Ga(\Bx)}{\Md x_i}.
\eeq{5.8a}
This can be rewritten more concisely as
\beq \BE(\Bx)\BL(\Bx)\Grad\BV(\Bx)+\BE(\Bx)\BD(\Bx)\BV(\Bx)-\BE(\Bx)\Bt(\Bx)=\BE(\Bx)\BR^\perp\Grad\BW(\Bx),
\eeq{5.9}
or alternatively as
\beq \widetilde{\BB}(\Bx)\cdot\Grad\Bu(\Bx)
+\widetilde{\BA}(\Bx)\Bu(\Bx)=\widetilde{\Br}(\Bx), \eeq{5.10}
where
\beq \widetilde{\BB}(\Bx)=\bpm \BE(\Bx)\BL(\Bx) \\ -\BE(\Bx)\BR^\perp \epm,\quad 
\widetilde{\BA}(\Bx)=\bpm \BE(\Bx)\BD(\Bx) \\ 0 \epm,\quad \widetilde{\Br}(\Bx)=\BE(\Bx)\Bt(\Bx),\quad
\Bu(\Bx)=\bpm \BV(\Bx)\\ \BW(\Bx) \epm.
\eeq{5.11}
In this form the analysis of the previous sections can be applied.


\section*{Acknowledgements}
The author thanks the National Science Foundation for support through grant DMS-1211359. The author is also
grateful to Ciprian Borcea, Claude Boutin, Alexander Kildishev, Sofia Mogilevskaya, Vladimir Shalaev, Valery Smyshlyaev, and  Sergei Tretyakov
for pointers to important references, either on their websites, or in person. Alexander and Natasha Movchan are thanked for comments on the manuscript.
The author is grateful to the Institute for Mathematics and its Applications at the University of Minnesota, where the final stages of the work were completed.

\end{document}